\pdfoutput=1

\documentclass[12pt,a4paper]{article}

\usepackage{ifthen} 
\newboolean{pdflatex}
\setboolean{pdflatex}{false} 

\newboolean{articletitles}
\setboolean{articletitles}{true} 

\newboolean{uprightparticles}
\setboolean{uprightparticles}{false} 

\usepackage{microtype}
\usepackage{lineno}  
\usepackage{xspace} 

\usepackage{graphicx}  
\usepackage{color}
\usepackage{colortbl}
\graphicspath{{./figs/}} 

\usepackage{rotating}
\usepackage{subfigure}
\usepackage{amsmath} 
\usepackage{amssymb}
\usepackage{amsfonts}
\usepackage{upgreek} 

\newcommand*\patchAmsMathEnvironmentForLineno[1]{%
\expandafter\let\csname old#1\expandafter\endcsname\csname #1\endcsname
\expandafter\let\csname oldend#1\expandafter\endcsname\csname
end#1\endcsname
 \renewenvironment{#1}%
   {\linenomath\csname old#1\endcsname}%
   {\csname oldend#1\endcsname\endlinenomath}%
}
\newcommand*\patchBothAmsMathEnvironmentsForLineno[1]{%
  \patchAmsMathEnvironmentForLineno{#1}%
  \patchAmsMathEnvironmentForLineno{#1*}%
}
\AtBeginDocument{%
\patchBothAmsMathEnvironmentsForLineno{equation}%
\patchBothAmsMathEnvironmentsForLineno{align}%
\patchBothAmsMathEnvironmentsForLineno{flalign}%
\patchBothAmsMathEnvironmentsForLineno{alignat}%
\patchBothAmsMathEnvironmentsForLineno{gather}%
\patchBothAmsMathEnvironmentsForLineno{multline}%
}

\usepackage{hyperref}    
\usepackage[all]{hypcap} 

\def\lhcb {\mbox{LHCb}\xspace}
\def\ux85 {\mbox{UX85}\xspace}

\def\cdf    {\mbox{CDF}\xspace}

\ifthenelse{\boolean{uprightparticles}}%
{

 \def\Pmu         {\ensuremath{\upmu}\xspace}

 \def\Ppi         {\ensuremath{\uppi}\xspace}

 \def\Ppsi        {\ensuremath{\uppsi}\xspace}

 \def\PDelta      {\ensuremath{\Delta}\xspace}                 
 \def\PXi      {\ensuremath{\Xi}\xspace}                 
 \def\PLambda      {\ensuremath{\Lambda}\xspace}                 
 \def\PSigma      {\ensuremath{\Sigma}\xspace}                 
 \def\POmega      {\ensuremath{\Omega}\xspace}                 
 \def\PUpsilon      {\ensuremath{\Upsilon}\xspace}                 
 
 \def\PB      {\ensuremath{\mathrm{B}}\xspace}                 
                  
 \def\PD      {\ensuremath{\mathrm{D}}\xspace}

 \def\PJ      {\ensuremath{\mathrm{J}}\xspace}                 
 \def\PK      {\ensuremath{\mathrm{K}}\xspace}

 \def\Pb      {\ensuremath{\mathrm{b}}\xspace}                 
 \def\Pc      {\ensuremath{\mathrm{c}}\xspace}

 \def\Pi      {\ensuremath{\mathrm{i}}\xspace}

}
{

 \def\Pmu         {\ensuremath{\mu}\xspace}

 \def\Ppi         {\ensuremath{\pi}\xspace}

 \def\Ppsi        {\ensuremath{\psi}\xspace}                 
                  
 \mathchardef\PDelta="7101
 \mathchardef\PXi="7104
 \mathchardef\PLambda="7103
 \mathchardef\PSigma="7106
 \mathchardef\POmega="710A
 \mathchardef\PUpsilon="7107
                  
 \def\PB      {\ensuremath{B}\xspace}                 
                  
 \def\PD      {\ensuremath{D}\xspace}

 \def\PJ      {\ensuremath{J}\xspace}                 
 \def\PK      {\ensuremath{K}\xspace}

 \def\Pb      {\ensuremath{b}\xspace}                 
 \def\Pc      {\ensuremath{c}\xspace}

 \def\Pi      {\ensuremath{i}\xspace}

}

\def\mup        {\ensuremath{\Pmu^+}\xspace}
\def\mun        {\ensuremath{\Pmu^-}\xspace} 

\def\cquark    {\ensuremath{\Pc}\xspace}
\def\cquarkbar {\ensuremath{\overline \cquark}\xspace}

\def\bquark    {\ensuremath{\Pb}\xspace}

\def\pion  {\ensuremath{\Ppi}\xspace}

\def\pip   {\ensuremath{\pion^+}\xspace}
\def\pim   {\ensuremath{\pion^-}\xspace}

\def\kaon  {\ensuremath{\PK}\xspace}
  \def\Kbar  {\kern 0.2em\overline{\kern -0.2em \PK}{}\xspace}

\def\Kz    {\ensuremath{\kaon^0}\xspace}
\def\Kzb   {\ensuremath{\Kbar^0}\xspace}
\def\KzKzb {\ensuremath{\Kz \kern -0.16em \Kzb}\xspace}
\def\Kp    {\ensuremath{\kaon^+}\xspace}
\def\Km    {\ensuremath{\kaon^-}\xspace}

\def\KpKm  {\ensuremath{\Kp \kern -0.16em \Km}\xspace}

  \def\Dbar    {\kern 0.2em\overline{\kern -0.2em \PD}{}\xspace}
\def\D       {\ensuremath{\PD}\xspace}

\def\Dz      {\ensuremath{\D^0}\xspace}
\def\Dzb     {\ensuremath{\Dbar^0}\xspace}
\def\DzDzb   {\ensuremath{\Dz {\kern -0.16em \Dzb}}\xspace}
\def\Dp      {\ensuremath{\D^+}\xspace}
\def\Dm      {\ensuremath{\D^-}\xspace}

\def\DpDm    {\ensuremath{\Dp {\kern -0.16em \Dm}}\xspace}

\def\B       {\ensuremath{\PB}\xspace}
\def\Bbar    {\ensuremath{\kern 0.18em\overline{\kern -0.18em \PB}{}}\xspace}

\def\Bc      {\ensuremath{\B_\cquark^+}\xspace}
\def\Bcp     {\ensuremath{\B_\cquark^+}\xspace}

\def\jpsi     {\ensuremath{{\PJ\mskip -3mu/\mskip -2mu\Ppsi\mskip 2mu}}\xspace}
\def\psitwos  {\ensuremath{\Ppsi{(2S)}}\xspace}

  \def\Y#1S{\ensuremath{\PUpsilon{(#1S)}}\xspace}

\def\Lbar {\ensuremath{\kern 0.1em\overline{\kern -0.1em\PLambda}}\xspace}

\def\BF         {{\ensuremath{\cal B}\xspace}}

\def\BR         {\BF}
\newcommand{\decay}[2]{\ensuremath{#1\!\to #2}\xspace}         

\def\to                 {\ensuremath{\rightarrow}\xspace}

\def\AT#1     {\ensuremath{A_{\mathrm{T}}^{#1}}\xspace}           

\def\C#1      {\ensuremath{\mathcal{C}_{#1}}\xspace}                       
\def\Cp#1     {\ensuremath{\mathcal{C}_{#1}^{'}}\xspace}                    
\def\Ceff#1   {\ensuremath{\mathcal{C}_{#1}^{\mathrm{(eff)}}}\xspace}        
\def\Cpeff#1  {\ensuremath{\mathcal{C}_{#1}^{'\mathrm{(eff)}}}\xspace}       
\def\Ope#1    {\ensuremath{\mathcal{O}_{#1}}\xspace}                       
\def\Opep#1   {\ensuremath{\mathcal{O}_{#1}^{'}}\xspace}                    

\newcommand{\tev}{\ensuremath{\mathrm{\,Te\kern -0.1em V}}\xspace}
\newcommand{\gev}{\ensuremath{\mathrm{\,Ge\kern -0.1em V}}\xspace}
\newcommand{\mev}{\ensuremath{\mathrm{\,Me\kern -0.1em V}}\xspace}
\newcommand{\kev}{\ensuremath{\mathrm{\,ke\kern -0.1em V}}\xspace}
\newcommand{\ev}{\ensuremath{\mathrm{\,e\kern -0.1em V}}\xspace}
\newcommand{\gevc}{\ensuremath{{\mathrm{\,Ge\kern -0.1em V\!/}c}}\xspace}
\newcommand{\mevc}{\ensuremath{{\mathrm{\,Me\kern -0.1em V\!/}c}}\xspace}
\newcommand{\gevcc}{\ensuremath{{\mathrm{\,Ge\kern -0.1em V\!/}c^2}}\xspace}
\newcommand{\gevgevcccc}{\ensuremath{{\mathrm{\,Ge\kern -0.1em V^2\!/}c^4}}\xspace}
\newcommand{\mevcc}{\ensuremath{{\mathrm{\,Me\kern -0.1em V\!/}c^2}}\xspace}

\def\mum  {\ensuremath{\,\upmu\rm m}\xspace}

\def\mub{\ensuremath{\rm \,\upmu b}\xspace}

\def\invfb   {\ensuremath{\mbox{\,fb}^{-1}}\xspace}

\def\mhz  {\ensuremath{{\rm \,MHz}}\xspace}
\def\khz  {\ensuremath{{\rm \,kHz}}\xspace}

\newcommand{\stat}{\ensuremath{\mathrm{(stat)}}\xspace}
\newcommand{\syst}{\ensuremath{\mathrm{(syst)}}\xspace}

\def\gsim{{~\raise.15em\hbox{$>$}\kern-.85em
          \lower.35em\hbox{$\sim$}~}\xspace}
\def\lsim{{~\raise.15em\hbox{$<$}\kern-.85em
          \lower.35em\hbox{$\sim$}~}\xspace}

\def\sPlot{\mbox{\em sPlot}}

\def\pt         {\mbox{$p_{\rm T}$}\xspace}

\def\evtgen     {\mbox{\textsc{EvtGen}}\xspace}
\def\pythia     {\mbox{\textsc{Pythia}}\xspace}

\def\geant      {\mbox{\textsc{Geant4}}\xspace}

\def\photos     {\mbox{\textsc{Photos}}\xspace}

\def\tell1  {TELL1\xspace}
\def\ukl1   {UKL1\xspace}

\usepackage{cite} 
\usepackage{mciteplus}

\usepackage{longtable} 

\newcommand{\BcToPsipi}{\decay{\Bcp}{\psitwos\pip}}
\newcommand{\BcToJpsipi}{\decay{\Bcp}{\jpsi\pip}}
\newcommand{\BcToJpsiK}{\decay{\Bcp}{\jpsi\Kp}}
\newcommand{\JpsiToMumu}{\decay{\jpsi}{\mup\mun}}
\newcommand{\PsiToMumu}{\decay{\psitwos}{\mup\mun}}

\begin{document}

\renewcommand{\thefootnote}{\fnsymbol{footnote}}
\setcounter{footnote}{1}


\begin{titlepage}
\pagenumbering{roman}

\vspace*{-1.5cm}
\centerline{\large EUROPEAN ORGANIZATION FOR NUCLEAR RESEARCH (CERN)}
\vspace*{1.5cm}
\hspace*{-0.5cm}
\begin{tabular*}{\linewidth}{lc@{\extracolsep{\fill}}r}
\ifthenelse{\boolean{pdflatex}}
{\vspace*{-2.7cm}\mbox{\!\!\!\includegraphics[width=.14\textwidth]{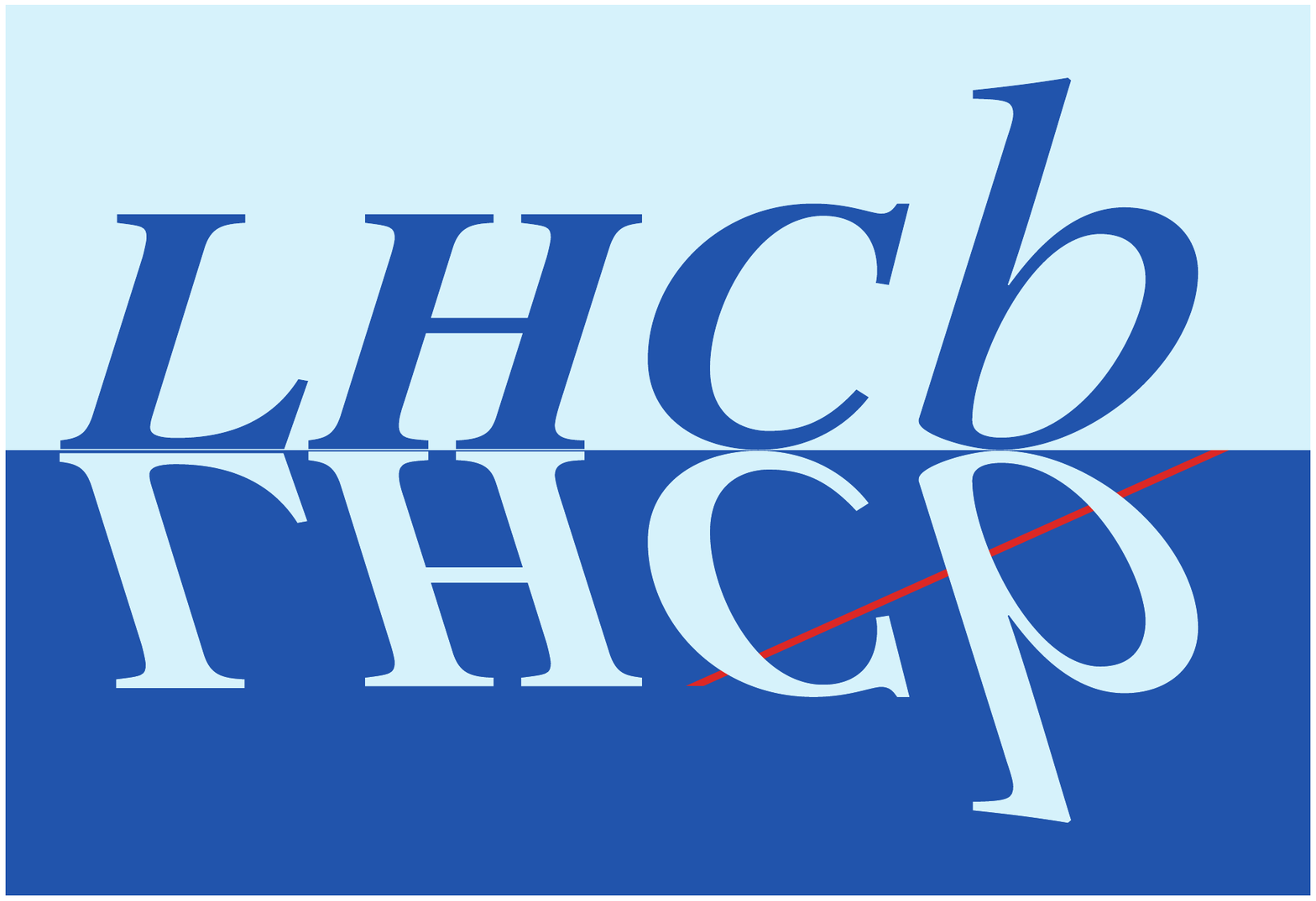}} & &}%
{\vspace*{-1.2cm}\mbox{\!\!\!\includegraphics[width=.12\textwidth]{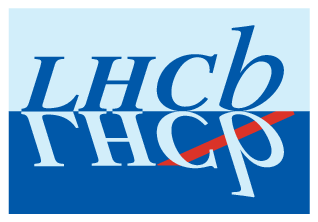}} & &}%
\\
 & & CERN-PH-EP-2013-028 \\  
 & & LHCb-PAPER-2012-054 \\  
 & & 6 March 2013 \\
\end{tabular*}

\vspace*{4.0cm}

{\bf\boldmath\huge
\begin{center}
  Observation of the decay $B_{c}^{+} \rightarrow \psi(2S)\pi^{+}$ 
\end{center}
}

\vspace*{2.0cm}

\begin{center}
The LHCb collaboration\footnote{Authors are listed on the following pages.}
\end{center}

\vspace{\fill}

\begin{abstract}
  \noindent
  The decay $\BcToPsipi$ with $\PsiToMumu$ is observed with a significance of $5.2\,\sigma$ using $pp$ collision data corresponding to an integrated luminosity of $1.0\,\invfb$ collected by the LHCb experiment. The branching fraction of $\BcToPsipi$ decays relative to that of the $\BcToJpsipi$ mode is measured to be 
\begin{equation*}
\frac{\BR(\BcToPsipi)}{\BR(\BcToJpsipi)}
= 0.250 \pm 0.068 \,\stat \pm 0.014 \,\syst \pm 0.006 \,(\BR).
\end{equation*}
The last term is the uncertainty on the ratio $\BR(\PsiToMumu)/\BR(\JpsiToMumu)$. 
\end{abstract}

\vspace*{2.0cm}

\begin{center}
Submitted to Physical Review Letters
\end{center}

\vspace{\fill}

{\footnotesize 
\centerline{\copyright~CERN on behalf of the \lhcb collaboration, license \href{http://creativecommons.org/licenses/by/3.0/}{CC-BY-3.0}.}}
\vspace*{2mm}

\end{titlepage}


\newpage
\setcounter{page}{2}
\mbox{~}
\newpage

\centerline{\large\bf LHCb collaboration}
\begin{flushleft}
\small
R.~Aaij$^{40}$, 
C.~Abellan~Beteta$^{35,n}$, 
B.~Adeva$^{36}$, 
M.~Adinolfi$^{45}$, 
C.~Adrover$^{6}$, 
A.~Affolder$^{51}$, 
Z.~Ajaltouni$^{5}$, 
J.~Albrecht$^{9}$, 
F.~Alessio$^{37}$, 
M.~Alexander$^{50}$
% ...

{\footnotesize \it
% ...
$ ^{5}$Clermont Universit\'{e}, Universit\'{e} Blaise Pascal, CNRS/IN2P3, LPC, Clermont-Ferrand, France\\
$ ^{6}$CPPM, Aix-Marseille Universit\'{e}, CNRS/IN2P3, Marseille, France\\
% ...
$ ^{9}$Fakult\"{a}t Physik, Technische Universit\"{a}t Dortmund, Dortmund, Germany\\
% ...
$ ^{35}$Universitat de Barcelona, Barcelona, Spain\\
$ ^{36}$Universidad de Santiago de Compostela, Santiago de Compostela, Spain\\
$ ^{37}$European Organization for Nuclear Research (CERN), Geneva, Switzerland\\
% ...
$ ^{40}$Nikhef National Institute for Subatomic Physics, Amsterdam, The Netherlands\\
% ...
$ ^{45}$H.H. Wills Physics Laboratory, University of Bristol, Bristol, United Kingdom\\
% ...
$ ^{50}$School of Physics and Astronomy, University of Glasgow, Glasgow, United Kingdom\\
$ ^{51}$Oliver Lodge Laboratory, University of Liverpool, Liverpool, United Kingdom\\
% ...
$ ^{n}$LIFAELS, La Salle, Universitat Ramon Llull, Barcelona, Spain\\
% ...
}
\end{flushleft}

\cleardoublepage


\renewcommand{\thefootnote}{\arabic{footnote}}
\setcounter{footnote}{0}


\pagestyle{plain} 
\setcounter{page}{1}
\pagenumbering{arabic}


\noindent The $\Bc$ meson, discovered by $\cdf$ at the Tevatron~\cite{Abe:1998wi}, is the only known meson composed of two flavours of heavy quarks, charm and beauty. Both quarks can decay via the weak interaction with the other quark being considered as a spectator, therefore a wide range of decay channels are possible. However, only a few of these channels have been experimentally observed~\cite{Abe:1998wi,Abulencia:2006zu,Aaltonen:2007gv,Abazov:2008kv}.
The LHC opens a new era for $\Bc$ physics, with an expected production cross-section of $\sim 0.4\,\mub$ at centre-of-mass energy $\sqrt{s} = 7$ TeV for the $\Bc$ meson~\cite{Chang:2003cr,0256-307X-27-6-061302}. 
The LHCb experiment has observed the decay $\BcToJpsipi$~\cite{PhysRevLett.109.232001}, and new channels such as $\Bc \to \jpsi\pip\pip\pim$~\cite{LHCb:2012ag} have started to emerge. 

We report here the first observation of the decay $\BcToPsipi$ with $\PsiToMumu$ and the measurement of the ratio of branching fractions $\BR(\BcToPsipi)/\BR(\BcToJpsipi)$. 
The inclusion of charge conjugate modes is implied throughout the paper. 
The relativistic quark model~\cite{Ebert:2003cn} and several other models~\cite{PhysRevD.49.3399,PhysRevD.56.4133,Colangelo:1999zn,Qiao:2012hp} make various theoretical predictions for this ratio of branching fractions.
As a two-body decay, $\BcToPsipi$ is under better control theoretically than $\Bc \rightarrow J/\psi \pi^{+}\pi^{-}\pi^{+}$, and therefore this measurement is particularly useful to test the models of $\Bc$ decays.
The $\BcToJpsipi$ decay mode is chosen as the normalisation channel because of its identical final state and similar event topology. Both channels take advantage of the large trigger efficiency due to the two muons in the final state.

The analysis is based on $pp$ collision data corresponding to an integrated luminosity of $1.0\,\invfb$ at $\sqrt{s} = 7 \tev$ collected with the LHCb detector in 2011. 
The detector~\cite{Alves:2008zz} is a single-arm forward
spectrometer covering the \mbox{pseudorapidity} range $2< \eta <5$,
designed for the study of particles containing \bquark or \cquark
quarks. The detector includes a high precision tracking system
consisting of a silicon-strip vertex detector surrounding the $pp$
interaction region, a large-area silicon-strip detector located
upstream of a dipole magnet with a bending power of about
$4{\rm\,Tm}$, and three stations of silicon-strip detectors and straw
drift tubes placed downstream. The combined tracking system has
momentum resolution $\Delta p/p$ that varies from 0.4\% at 5\,\gevc to
0.6\% at 100\,\gevc, and impact parameter (IP) resolution of 20\,\mum for
tracks with high transverse momentum ($p_\text{T}$). 
Charged hadrons are identified using two ring-imaging Cherenkov detectors and good kaon-pion separation is achieved for tracks with momentum between 5\,\gevc and 100\,\gevc. 
Photon, electron and hadron candidates are identified by a calorimeter system consisting of
scintillating-pad and preshower detectors, an electromagnetic
calorimeter and a hadronic calorimeter. Muons are identified by a
system composed of alternating layers of iron and multiwire
proportional chambers. The trigger system~\cite{Aaij:2012me} consists of a
hardware stage, based on information from the calorimeter and muon
systems, followed by a software trigger that applies a full event
reconstruction and reduces the event rate from $1\,\mhz$ to around $3\,\khz$.

Candidate $\Bc \rightarrow \psi \pi^{+}$ decays with $\psi \rightarrow \mup\mun$, where $\psi$ denotes $\jpsi$ or $\psitwos$, are selected by requiring a single muon or dimuon with high $\pt$ in the hardware trigger. In the software trigger, a charged particle is required to have $\pt > 1.7\,\gevc$, or $\pt > 1\,\gevc$ if identified as a muon; alternatively a dimuon trigger requires two oppositely charged muons with $\pt > 500\,\mevc$, the invariant mass of the muon pair $M_{\mup\mun} > 2.95\,\gevcc$, and that the muon track pair has a decay length significance with respect to the primary vertex greater than 5.

Further offline selections require both muons to have $\pt > 550\,\mevc$, and a track fit $\chi^2_{\text{tr}}$ per degree of freedom ($\chi^2_{\text{tr}}/\text{ndf}$) of less than 5. The mass of the $\psi$ candidate is required to be within a window of $100\,\mevcc$ centred around the known $\psi$ mass ($3686\,\mevcc$ for $\psitwos$ and $3097\,\mevcc$ for $\jpsi$)~\cite{PDG2012}. The $\psi$ vertex fit $\chi^2_{\text{vtx}}/\text{ndf}$ is required to be less than 20, and the $\psi$ decay length significance larger than 5.

The $\Bc$ candidate is reconstructed from the $\psi$ and a bachelor pion.
The pion is required to have $p_{\text{T}} > 500\,\mevc$, a track fit $\chi^2_{\text{tr}}/\text{ndf} < 10$ and IP $\chi^2_{\text{IP}}$ with respect to the primary interaction great than 4. The IP $\chi^2_{\text{IP}}$ is defined as the difference between the $\chi^2$ of the primary vertex reconstructed with and without the considered track. The $\Bc$ candidate is required to have mass within $0.5\,\gevcc$ around the world average value~\cite{PDG2012} and a vertex fit $\chi^2_{\text{vtx}}/\text{ndf} < 16$.

A boosted decision tree (BDT)~\cite{Breiman}, trained on data and simulation, is used to perform further background suppression.
The $pp$ collisions are simulated using \pythia~6.4~\cite{Sjostrand:2006za} with a specific \lhcb configuration~\cite{LHCb-PROC-2010-056}. The $\Bc$ mesons are generated through the dominant hard subprocess $gg \to \Bc + \bquark + \cquarkbar$ with the dedicated generator \textsc{Bcvegpy}~\cite{Chang:2003cq,Chang:2005hq}. Decays of hadronic particles are described by \evtgen~\cite{Lange:2001uf} in which final state radiation is generated using \photos~\cite{Golonka:2005pn}. The interaction of the generated particles with the detector and its
response are implemented using \geant~\cite{Allison:2006ve, *Agostinelli:2002hh} as described in Ref.~\cite{LHCb-PROC-2011-006}.

The choice of the variables used to train the BDT is based on two considerations: their power to separate signal and background, and the similarity of the distributions for the $\BcToJpsipi$ and $\BcToPsipi$ candidates that causes the systematic uncertainties in the selections to cancel when the ratio of branching fractions is determined. The BDT input variables are:
the $\pip$ IP $\chi^2_{\text{IP}}$;
the $\Bc$ vertex fit $\chi^2_{\text{vtx}}/\text{ndf}$;
the $\Bc$ IP $\chi^2_{\text{IP}}$;
the $\chi^2$ of the distance between the $\Bc$ vertex and the associated primary vertex; 
the $p_{\text{T}}$ of the $\Bc$ candidate;
and the $\chi^2$ from a refit of the $\Bc$ decay vertex~\cite{Hulsbergen:2005pu} using a $\jpsi$ or $\psitwos$ mass constraint and a constraint that the $\Bc$ candidate points to the primary vertex.

The BDT is trained using a $\BcToJpsipi$ simulation sample for the signal and sidebands from the $\BcToJpsipi$ mass spectrum ($6164 < M_{\jpsi \pi} < 6206\,\mevcc$ or $6346 < M_{\jpsi \pi} < 6388\,\mevcc$) for the background. 
The trained BDT is then applied to the data, and a signal estimator is calculated for each candidate; a large value indicates a signal-like candidate. The cut on the estimator is optimised to maximise the $\BcToPsipi$ signal significance. The BDT selection efficiencies, estimated from simulation, for $\BcToPsipi$ and $\BcToJpsipi$ candidates are 35.8\% and 37.2\% respectively, and the fraction of accepted background is $4.8 \times 10^{-4}$ as estimated from the sideband data.

\begin{figure}[th]
\begin{center}
\includegraphics[width=0.48\textwidth]{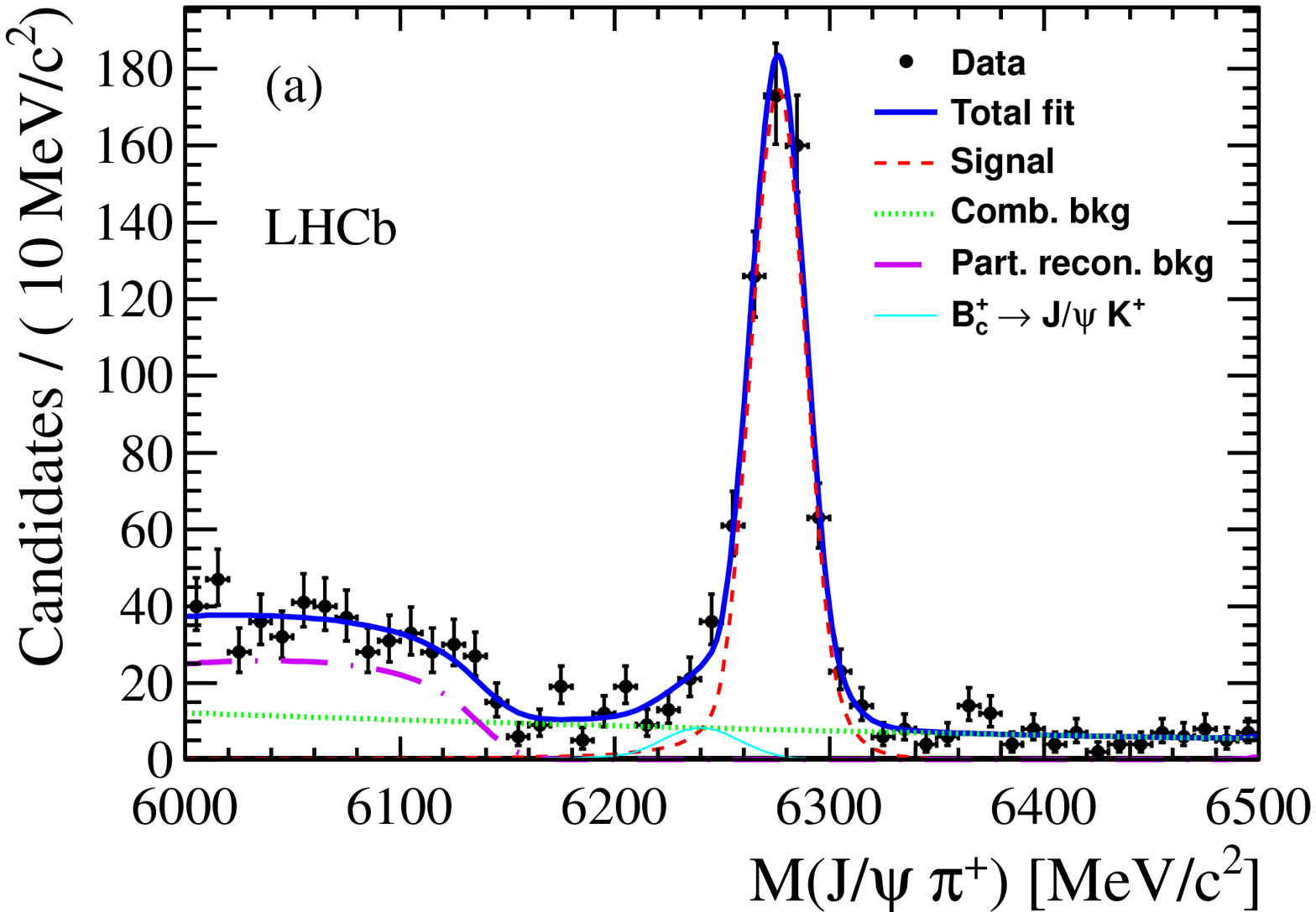}
\includegraphics[width=0.48\textwidth]{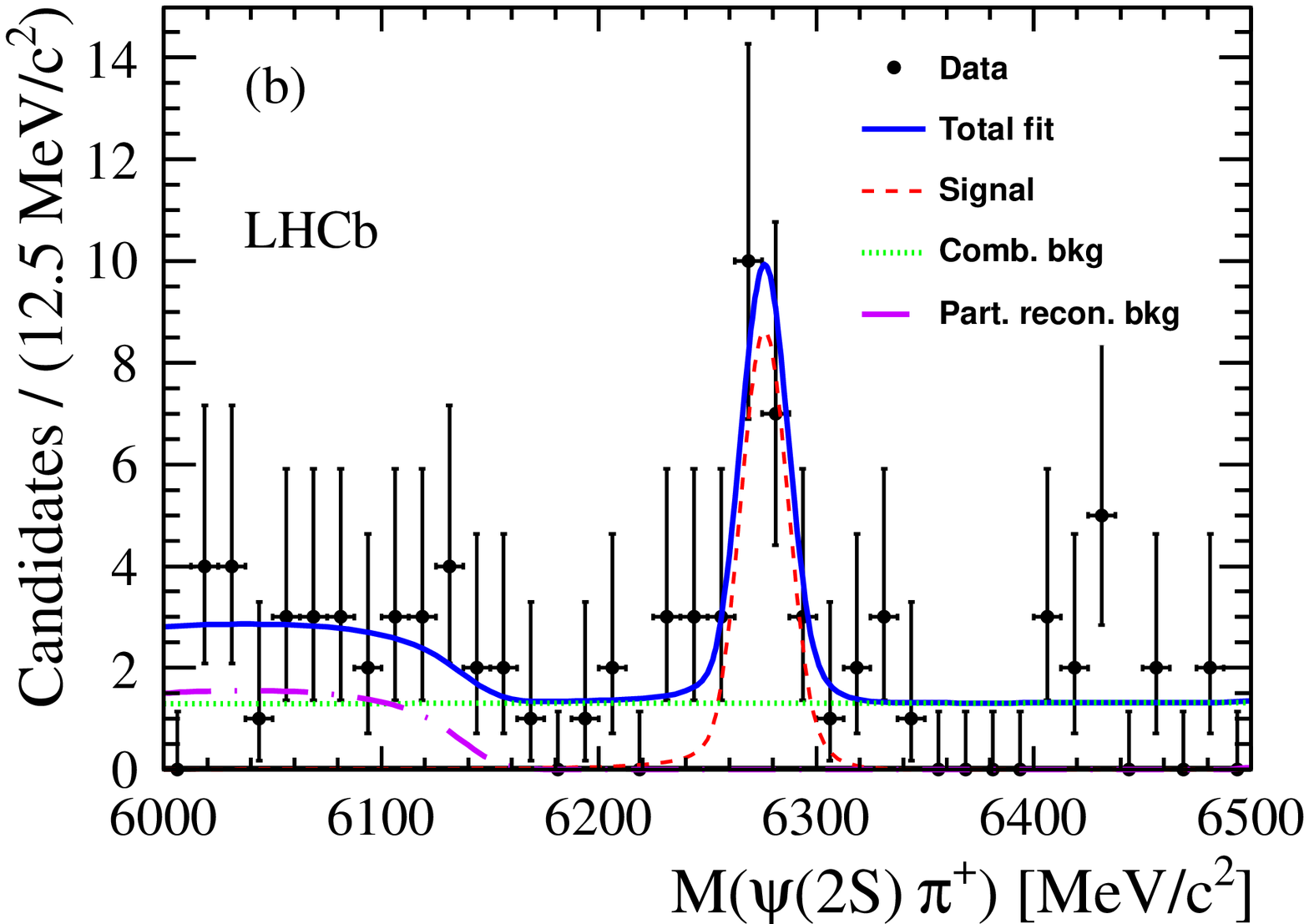}
\end{center}
\caption{Invariant mass distributions of $\Bc$ candidates reconstructed as 
(a) $\BcToJpsipi$ and 
(b) $\BcToPsipi$. 
Points with error bars (black) show the data, the thick solid line (blue) represents the fit of the data, the dashed line (red) the signal distribution, the dotted line (green) the combinatorial background, the dot-dashed line (purple) the partially reconstructed background, and the thin solid line (light blue) the background from the $\BcToJpsiK$ channel.}
\label{fig:fitting-result}
\end{figure}

After the BDT selection, it is further required that the unconstrained dimuon invariant mass is in the range $3030 < M_{\mu\mu} < 3170\,\mevcc$ for $\jpsi$ and $3620 < M_{\mu\mu} < 3760\,\mevcc$ for $\psitwos$. Information on particle identification for pions and kaons is also used to suppress the reflection background due to $\BcToJpsiK$ decays. Figure \ref{fig:fitting-result} shows the invariant mass distributions of the $\BcToJpsipi$ and $\BcToPsipi$ candidates.

The relative branching fraction is calculated using
\begin{equation}
\frac{\BR(\BcToPsipi,\PsiToMumu)}{\BR(\BcToJpsipi,\JpsiToMumu)}
= \frac{N(\BcToPsipi)}{N(\BcToJpsipi)} 
\cdot \frac{\varepsilon(\BcToJpsipi)}{\varepsilon(\BcToPsipi)},
\label{eqn:fitting-brdef}
\end{equation}
where $N$ is the number of selected signal events and $\varepsilon$ is the total efficiency. 

The signal yields are obtained by performing an extended maximum likehood fit to the $\Bc$ mass spectra in Fig. \ref{fig:fitting-result}. The signal is modelled with a double-sided Crystal Ball function~\cite{Skwarnicki:1986xj} with the tail parameters on both sides determined from simulation. 
The main background component for both channels is combinatorial and is modelled using an exponential function. At the lower end of the mass spectrum, the contribution from the partially reconstructed background is modelled by an ARGUS function~\cite{Albrecht1990278} convolved with a Gaussian distribution. For the $\BcToJpsipi$ decay, the Cabibbo suppressed channel $\BcToJpsiK$ also contributes, and is fitted with a double-sided Crystal Ball function with all parameters fixed to values obtained from simulation.
The observed signal yields are $595 \pm 29$ for $\BcToJpsipi$ and $20 \pm 5$ for $\BcToPsipi$. Therefore the ratio of yields is
\begin{equation*}
  \frac{N(\BcToPsipi)}{N(\BcToJpsipi)} = 0.034 \pm 0.009 \,\stat.
\end{equation*}

The total efficiency is the product of the detector acceptance, and the trigger, reconstruction and selection efficiencies. Each contribution has been determined using simulated events for the two channels, and the ratio of the total efficiencies has been evaluated to be
\begin{equation*}
\frac{\varepsilon(\BcToPsipi)}{\varepsilon(\BcToJpsipi)} = 1.040 \pm 0.009,
\label{eqn:efficiency}
\end{equation*}
where the uncertainty is due to the limited size of the simulated sample.

Several sources of systematic uncertainty have been considered.
The measured ratio of signal yields is expected to be independent of the BDT selection, given that the distributions of training variables are very similar for the two channels. The ratio of signal yields is measured for different cuts on the BDT response, and is constant within the statistical uncertainties. The average of these ratios differs from the nominal value by 4.5\%,
which is taken as the systematic uncertainty due to the BDT selection.

The $\BcToPsipi$ signal is fitted with a double-sided Crystal Ball function. 
Alternatively we determine the signal shape directly from the simulation using kernel estimation~\cite{Cranmer2001198}, and convolve it with a Gaussian function to take into account the detector resolution while allowing the mean of the mass to vary. This results in a $1.7\%$ difference with respect to the nominal ratio, which is taken as the uncertainty due to the signal shape.

To consider the contribution from partially reconstructed background, the background is fitted with an exponential function within a narrower range ($6164 < M_{\psi\pi} < 6500\,\mevcc$). This results in 
a 2.9\% change with respect to the nominal fit, and is assigned as a systematic uncertainty.

The statistical uncertainty on the simulation when estimating the ratio of efficiencies leads to an uncertainty of 0.9\% on the ratio of branching fractions.
The difference between data and simulation introduces a systematic uncertainty, especially from variables used as input for the BDT. The distributions of these variables in simulation and data are compared, after the background is subtracted from the data using the $\sPlot$ technique~\cite{Pivk:2004ty}.
The difference is found to be negligible compared to the statistical fluctuation.

A summary of systematic uncertainties is given in Table \ref{tab:syst-sum}. The total systematic uncertainty is 5.7\%, with the most significant contribution coming from the BDT selection. Taking the systematic uncertainty into account and using the likelihood ratio test $\sqrt{-2\log(\mathcal{L}_{\text{B}}/\mathcal{L}_{\text{S+B}})}$~\cite{Cowan:2010js}, the significance of the $\BcToPsipi$ decay is estimated to be a $5.2\,\sigma$, where $\mathcal{L}_{\text{B}}$ and $\mathcal{L}_{\text{S+B}}$ represent the likelihood of the background-only hypothesis and the signal-plus-background hypothesis respectively.
\begin{table}[tb]
\caption{Relative systematic uncertainties.}
\begin{center}\begin{tabular}{lr}
Component & Value \,(\%) \\
\hline
BDT selection & 4.5 \\
Signal shape & 1.7 \\
Background shape & 2.9 \\
Simulation sample size & 0.9 \\
\hline
Total & 5.7 \\
\end{tabular}\end{center}
\label{tab:syst-sum}
\end{table}

In summary, a search for the decay $\BcToPsipi$ has been performed using a data sample corresponding to an integrated luminosity of $1.0\,\invfb$ at $\sqrt{s} = 7\,\tev$ collected by LHCb in 2011. The signal yield is $20 \pm 5$ candidates, making the first observation of this decay channel. Using $\BcToJpsipi$ as normalisation channel, the ratio of branching fractions is measured to be
\begin{equation*}
\frac{\BR(\BcToPsipi,\PsiToMumu)}{\BR(\BcToJpsipi,\JpsiToMumu)} 
= 0.033 \pm 0.009 \,\stat \pm 0.002 \,\syst.
\end{equation*}

Furthermore, taking $\BR(\jpsi\to e^+ e^-) = (5.94 \pm 0.06)\%$ and $\BR(\psitwos \to e^+ e^-) = (7.73 \pm 0.17) \times 10^{-3}$ ~\cite{PDG2012} and assuming universality of the electroweak interaction, we obtain
\begin{equation*}
\frac{\BR(\BcToPsipi)}{\BR(\BcToJpsipi)}
= 0.250 \pm 0.068 \,\stat \pm 0.014 \,\syst \pm 0.006 \,(\BR),
\end{equation*}
where the last term accounts for the uncertainty on $\BR(\PsiToMumu)/\BR(\JpsiToMumu)$. 
This result favours the prediction made by the relativistic quark model~\cite{Ebert:2003cn} in comparison with the other models.

\section*{Acknowledgements}
\noindent We thank Prof.\,Chao-Hsi Chang for valuable discussions.
We express our gratitude to our colleagues in the CERN
accelerator departments for the excellent performance of the LHC. We
thank the technical and administrative staff at the LHCb
institutes. We acknowledge support from CERN and from the national
agencies: CAPES, CNPq, FAPERJ and FINEP (Brazil); NSFC (China);
CNRS/IN2P3 and Region Auvergne (France); BMBF, DFG, HGF and MPG
(Germany); SFI (Ireland); INFN (Italy); FOM and NWO (The Netherlands);
SCSR (Poland); ANCS/IFA (Romania); MinES, Rosatom, RFBR and NRC
``Kurchatov Institute'' (Russia); MinECo, XuntaGal and GENCAT (Spain);
SNSF and SER (Switzerland); NAS Ukraine (Ukraine); STFC (United
Kingdom); NSF (USA). We also acknowledge the support received from the
ERC under FP7. The Tier1 computing centres are supported by IN2P3
(France), KIT and BMBF (Germany), INFN (Italy), NWO and SURF (The
Netherlands), PIC (Spain), GridPP (United Kingdom). We are thankful
for the computing resources put at our disposal by Yandex LLC
(Russia), as well as to the communities behind the multiple open
source software packages that we depend on.

\addcontentsline{toc}{section}{References}
\bibliographystyle{LHCb}
\bibliography{main}


\end{document}